\documentclass{article}
\usepackage{spconf,amsmath,graphicx,hyperref}
\usepackage{algorithm}
\usepackage{algpseudocode}
\usepackage{float}
\usepackage{booktabs}
\usepackage{multirow}
\usepackage{subcaption}

\ninept
\title{Listening, Imagining \& Refining: A Heuristic Optimized ASR Correction Framework with LLMs}
\name{Yutong Liu$^{1}$, Ziyue Zhang$^{1}$, Cheng Huang$^{3}$, Yongbin Yu$^{1,\star}$, Xiangxiang Wang$^{1,\star}$, Yuqing Cai$^{1}$, Nyima Tashi$^{1,2}$}
\address{$^{1}$ School of Information and Software Engineering, \\ 
University of Electronic Science and Technology of China, Chengdu, China. \\
$^{2}$ School of Information Science and Technology, Tibet University, Lhasa, China.\\
$^{3}$ Department of Ophthalmology, University of Texas Southwestern Medical Center, USA\\
ybyu@uestc.edu.cn, xxwang@uestc.edu.cn}

\begin{document}
%
\maketitle
\begin{abstract}
Automatic Speech Recognition (ASR) systems remain prone to errors that affect downstream applications. In this paper, we propose LIR-ASR, a heuristic optimized iterative correction framework using LLMs, inspired by human auditory perception. LIR-ASR applies a "Listening-Imagining-Refining" strategy, generating phonetic variants and refining them in context. A heuristic optimization with finite state machine (FSM) is introduced to prevent the correction process from being trapped in local optima and rule-based constraints help  maintain semantic fidelity. Experiments on both English and Chinese ASR outputs show that LIR-ASR achieves average reductions in CER/WER of up to 1.5 percentage points compared to baselines, demonstrating substantial accuracy gains in transcription.
\end{abstract}


\section{Introduction}
\label{sec:intro}
Large audio model~\cite{radford2022whisper, zhang2023googleusm, meng2025dolphin, kashiwagi2025whale} have been widely adopted in automatic speech recognition (ASR) due to their superior robustness, generalization across languages, and ability to withstand diverse acoustic conditions. However, ASR outputs still suffer from low quality in many real-world scenarios, owing to environmental noise, overlapping speech, long-tail or out-of-vocabulary words, and varying speaker accents.

Several methods have been applied to ASR correction. For instance, Leng et al.~\cite{leng2022fastcorrect} introduced FastCorrect, a framework that utilizes a pre-trained language model to correct ASR errors efficiently. Building upon this, Leng et al.~\cite{leng2023softcorrect} proposed SoftCorrect, which incorporates a soft decision mechanism to improve correction flexibility. Wang et al.~\cite{Wang2020transformer} addressed domain-specific challenges by augmenting the Transformer model with entity retrieval capabilities, enhancing its ability to handle specialized terminology. Gu et al.~\cite{gu2024denoisinglm} developed DenoisingLM, a model that leverages denoising autoencoders to refine ASR outputs. However, these traditional approaches often exhibit limited generalization ability, struggle to handle diverse accents or acoustic conditions, and face significant challenges in supporting multiple languages effectively.

Large Language Models (LLMs)~\cite{grattafiori2024llama3, glm2024chatglm, yang2025qwen3, openai2024gpt4, deepseekai2025deepseekv3, kimiteam2025kimik2}, endowed with deep contextual understanding and generative prowess, offer a compelling solution for ASR post-processing. Recent efforts have demonstrated the efficacy of LLMs in error correction and transcription refinement. For instance, Ma et al.~\cite{ma2025asrerrorcorrectionusing} investigated ASR error correction using LLMs on N-best ASR outputs, with constrained decoding techniques and zero-shot application showing promising results. Similarly, Hu et al.~\cite{hu2024clozeger} introduced ClozeGER, a multimodal generative error correction approach using SpeechGPT, which significantly improved transcription fidelity across multiple standard datasets. Addressing ASR hallucinations further, Fang et al.~\cite{fang2025rllmcf} proposed a three-stage LLM-based framework—combining error pre-detection, chain-of-thought refinement, and reasoning-based verification—achieving significant relative reductions in CER/WER on AISHELL and LibriSpeech benchmarks. Additionally, Sachdev et al.~\cite{sachdev2024evolutionarypromptasr} explored evolutionary prompt design for LLM-based post-ASR error correction, enhancing the adaptability and performance of error correction models.

Despite advancements in ASR post-processing, existing methodologies face several limitations: 
1) Certain recognition errors appear contextually plausible within sentences, making them difficult for large language models to detect and correct. 
2) Recognition errors are often interdependent rather than isolated, where one mistake can reinforce or trigger others, complicating one-step correction. 
3) Correction models may generate linguistically plausible but semantically inconsistent substitutions, deviating from the intended speech content. These challenges collectively underscore the necessity for iterative refinement mechanisms like the one we propose, which can progressively resolve interdependent errors while maintaining semantic fidelity.

\begin{figure*}[t!]
  \includegraphics[width=\textwidth]{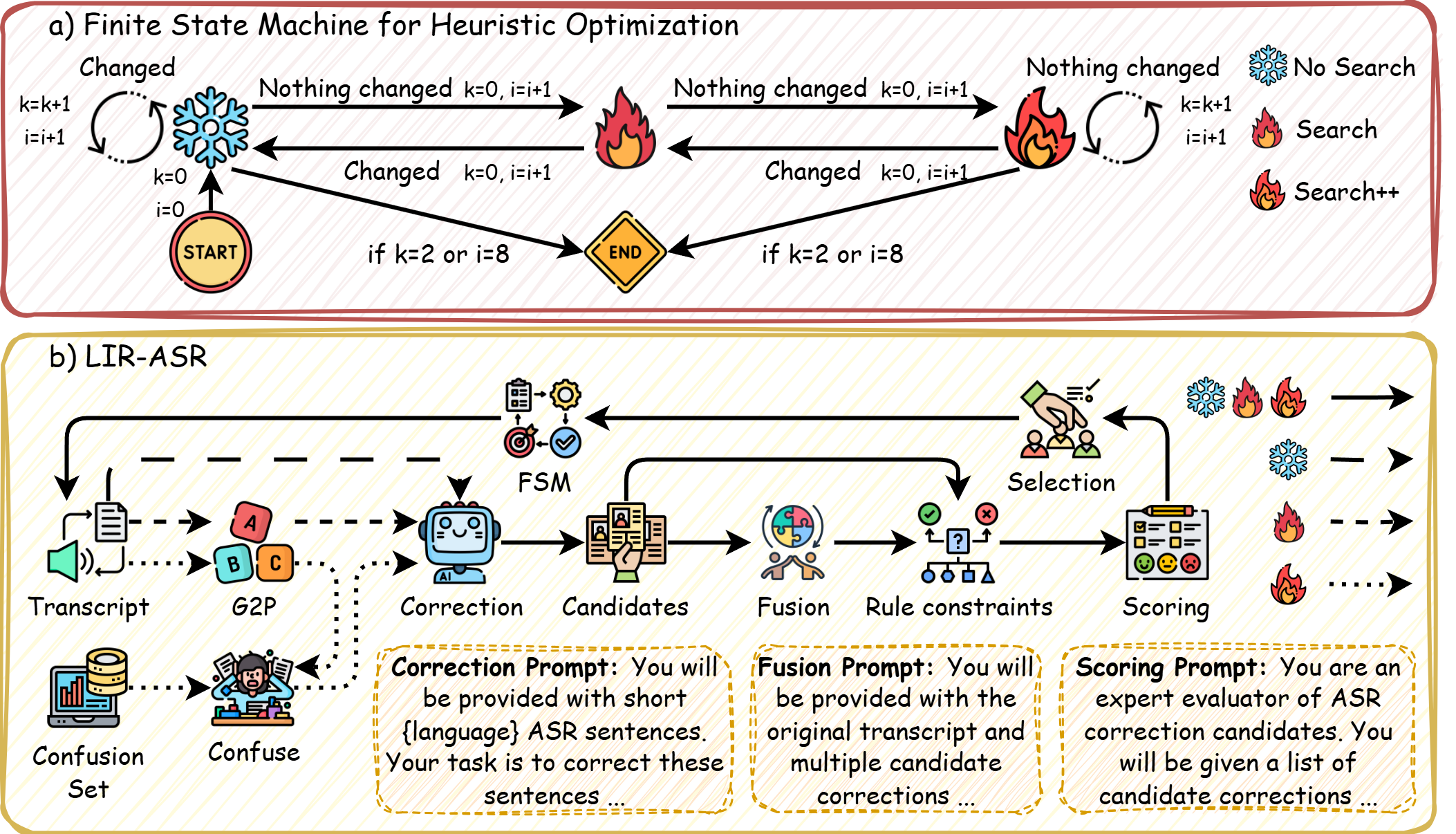}
  \caption{Overall framework of the proposed LIR-ASR.}
  \label{fig:LIR-ASR}
\end{figure*}

To bridge this gap, we propose \textbf{LIR-ASR}, a heuristic optimized ASR correction framework with LLMs. LIR-ASR draws inspiration from the human auditory process. When we suspect a mishearing, we often try to replace potentially erroneous sounds with phonetically similar alternatives and assess their contextual appropriateness. This mimics the "Listening-Imagining-Refining" (LIR) strategy: 1) Listening: Interpret the initial ASR output.2) Imagining: Generate plausible phonetic variants of uncertain words. 3) Refining: Evaluate these variants within the broader context to identify the most accurate transcription.
Incorporating heuristic optimization, the "Imagining" phase introduces controlled randomness, allowing the system to escape local optima and explore a wider solution space. This dynamic adjustment enhances the LLM's ability to produce more accurate transcriptions, especially in multilingual contexts.
Experiments on both English and Chinese ASR outputs demonstrate that LIR-ASR achieves average reductions in CER/WER of up to 1.5 percentage points compared to baselines, indicating substantial accuracy gains in transcription.
In summary, our primary contributions are:
\begin{itemize}
    \item A heuristic-optimized ASR correction framework, \textbf{LIR-ASR}, utilizing LLMs to effectively handle contextually plausible recognition errors.
    \item An iterative heuristic optimization strategy with a finite state machine (FSM), preventing the correction process from being trapped in local optima and allowing interdependent recognition errors to be progressively resolved.
    \item Rule-based constraints designed to guide the correction process, reducing the risk of linguistically plausible but semantically inconsistent substitutions introduced by LLMs.
\end{itemize}


\section{Methodology}\label{sec:method}
The LIR-ASR framework comprises two primary architectural components: a FSM that governs the neighbor search strategy and a heuristic optimization module for ASR correction, as illustrated in Fig.~\ref{fig:LIR-ASR}. The FSM-guided neighbor search implements the "Imagining" phase, systematically directing the candidate space exploration strategy. Concurrently, the iterative heuristic optimization module supports the "Refining" phase, ensuring that ASR outputs are progressively optimized through each iteration.

\subsection{Search Strategy Control via Finite State Machine}
To better regulate the neighbor searching strategy, we introduce a FSM that dynamically alternates among three states—No Search, Search, and Search++—as illustrated in Fig.~\ref{fig:LIR-ASR}(a). The process is initialized with the iteration counter set to $i=0$ and the consecutive no-change counter set to $k=0$. The FSM then proceeds as follows:

\textbf{No Search State:}
If the transcript improves (Changed), the system transitions to the Search state, resetting $k=0$ and incrementing the iteration counter ($i=i+1$). If no improvement occurs (Nothing changed), the FSM remains in the No Search State, incrementing both the no-change counter and the iteration counter ($k=k+1$, $i=i+1$).

\textbf{Search State:}
Upon improvement, the FSM moves to the Search++ State with $k=0$ and $i=i+1$. If no improvement is observed, it returns to the No Search State, resetting the counters ($k=0$, $i=i+1$).

\textbf{Search++ State:}
If a change occurs, the FSM transitions back to the Search State with $k=0$ and $i=i+1$. Otherwise, it remains in the Search++ State, incrementing the iteration counter while keeping the no-change counter at zero ($k=0$, $i=i+1$).

\textbf{Termination:}
The optimization terminates, entering the End State, if either the consecutive no-change counter reaches the predefined threshold ($k\geq2$) or the maximum number of iterations is exceeded ($i\geq8$).

This FSM-based design enables the euristic optimization process to systematically alternate between exploration and exploitation, maintaining a balance between global search and local refinement. The termination criteria prevent excessive iterations while avoiding stagnation in local optima.

\begin{table*}[t!]
\small
\centering
\caption{Comparison of ASR correction performance across different methods. Results are reported separately for Chinese (ZH) and English (EN). Bold numbers indicate the best performance within each column. $^*$ denotes the proposed method. The last column shows the average improvement in CER/WER over the uncorrected baseline.}
\begin{tabular}{l|l|cc|cc|c}
\toprule
& & \multicolumn{2}{c|}{\textbf{Whisper-medium}} 
& \multicolumn{2}{c|}{\textbf{Whisper-large-v3}} 
&  \\
\textbf{LLM} & \textbf{Method} & \textbf{ZH} & \textbf{EN} & \textbf{ZH} & \textbf{EN} & \textbf{$\Delta$CER / $\Delta$WER} \\
\midrule
none & none & 5.14 / 10.21 & 1.92 / 4.49 & 3.98 / 7.50 & 1.75 / 4.23 & -- \\
\midrule
\multirow{5}{*}{Qwen3-235B}
  & direct prompt       & 5.04 / 9.21  & 2.33 / 4.60  & 4.22 / 7.11  & 2.30 / 4.94  & -0.28 / +0.14 \\
  & evolutionary prompt & 5.11 / 8.98  & 2.51 / 5.09  & 5.64 / 9.11  & 2.96 / 5.54  & -0.86 / -0.57 \\
  & 3-best              & 4.80 / 9.50  & 1.93 / 4.35  & 3.94 / 7.40  & 1.94 / 4.38  & +0.04 / +0.20 \\
  & 6-best              & 4.54 / 9.13  & 1.90 / 4.36  & 3.88 / 7.29  & 1.99 / 4.57  & +0.12 / +0.27 \\
  & LIR-ASR$^*$         & \textbf{2.85 / 5.97} & \textbf{1.88 / 4.26} & \textbf{2.82 / 5.60} & \textbf{1.72 / 3.89} & \textbf{+0.88 / +1.68} \\
\midrule
\multirow{8}{*}{DeepSeek-V3.1}
  & direct prompt       & 3.59 / 5.95  & 1.99 / 4.38  & 3.19 / 6.41  & 2.18 / 4.38  & +0.46 / +1.33 \\
  & evolutionary prompt & 4.33 / 7.53  & 2.22 / 4.32  & 3.33 / 5.46  & 2.20 / 4.18  & +0.18 / +1.24 \\
  & 3-best              & 3.78 / 6.62  & 1.93 / 4.15  & 3.19 / 5.36  & 1.98 / 4.12  & +0.48 / +1.55 \\
  & 6-best              & 3.20 / 5.95  & 1.60 / 3.59  & 2.99 / 5.34  & 1.91 / 3.90  & +0.77 / +1.91 \\
  & LIR-ASR$^*$         & \textbf{3.20 / 5.91} & \textbf{1.68 / 3.70} & \textbf{2.89 / 5.23} & \textbf{1.59 / 3.60} & \textbf{+0.86 / +2.00} \\
  & \quad w/o rule      & 4.21 / 8.34  & 1.74 / 3.62  & 6.62 / 9.27  & 1.84 / 3.85  & -0.41 / +0.34 \\
  & \quad w/o multi-candidate & 3.89 / 7.50 & 1.73 / 3.79 & 3.24 / 6.40 & 1.73 / 3.90 & +0.55 / +1.21 \\
  & \quad w/o FSM       & 3.93 / 7.56  & 1.71 / 3.75  & 3.52 / 6.46  & 1.69 / 3.82  & +0.48 / +1.21 \\
  & \quad w/o search    & 3.80 / 7.43  & 1.68 / 3.70  & 3.25 / 5.98  & 1.71 / 3.86  & +0.59 / +1.37 \\
\bottomrule
\end{tabular}\label{tab:result}
\end{table*}

\subsection{Heuristic Optimization for ASR Correction}
The heuristic optimization process for ASR correction consists of five main components: neighbor generation ($\mathcal{N}$), correction ($\mathcal{C}$), candidate fusion ($\mathcal{F}$), rule constraints ($\mathcal{R}$), and scoring ($f$), collectively defining the optimization objective.

Let the transcript be $s$ with score $f(s)$, and let 
$S = \{s_0, s_1, \dots, s_n\}$ denote the set of candidate transcripts generated from the neighborhood $\mathcal{N}(s)$. Neighbor generation is performed via Grapheme-to-Phoneme (G2P) conversion and similar-sounding character substitutions, producing a richer set of candidate transcripts. The scoring component evaluates all valid candidates, and the LLM assigns both a score and a reasoning explanation for each candidate. Each candidate transcript is corrected in parallel via the LLM:
\begin{equation}
s'_i = \mathcal{C}(s_i), \quad \forall s_i \in \mathcal{N}(s),
\end{equation}
where $s'_i$ denotes the corrected candidate transcript. When multiple candidates exist, they are fused through an additional LLM step to produce a single transcript that optimally preserves semantic meaning while correcting phonetic or typographical errors:
\begin{equation}
s_\mathrm{fused} = \mathcal{F}(s, \{ s'_0, s'_1, \dots, s'_n \}),
\end{equation}
where $\mathcal{F}(\cdot)$ selects the candidate that is most semantically consistent, integrating the strengths of multiple corrected transcripts. Finally, the current transcript and the fused transcript are incorporated into the candidate set:
\begin{equation}
S' \gets \{ s'_0, s'_1, \dots, s'_n, s, s_\text{fused} \},
\end{equation}
ensuring that both the original and fused results are considered in the subsequent heuristic evaluation.

Rule constraints are subsequently employed to filter out unreliable candidates, as LLM-based corrections may introduce replacements that are linguistically plausible but semantically inconsistent. While neighbor searching effectively enhances candidate diversity, it inevitably introduces additional noise. There are two kinds of rule constraints: 1) \textbf{Phonetic consistency:} Candidate replacements must be sufficiently similar in pronunciation to the original words, measured by phonetic similarity metrics (e.g., Pinyin or G2P). \textbf{Length and structure consistency:} Candidates with excessive insertion or deletion are filtered out.

Let $s^{t}$ denote the transcript at iteration $t$, and let $f(s)$ be the corresponding score assigned by the LLM.  
At each iteration, the greedy acceptance rule selects the candidate $s'_i \in S'$ with the maximum score that is no worse than the current transcript:
\begin{equation}
s^{t+1} =
\begin{cases}
s'_i, & \text{if } f(s'_i) = \max_{s' \in S'} f(s') \ge f(s^{t}), \\
s^{t}, & \text{otherwise}.
\end{cases}
\end{equation}

Since the candidate set $S'$ at each step is finite and the acceptance rule ensures
\begin{equation}
f(s^{t+1}) \ge f(s^{t}),
\end{equation}
the sequence of scores $\{ f(s^{t}) \}$ is monotonically non-decreasing. Furthermore, because $f(s)$ is upper-bounded by a maximum achievable score $f_\text{max}$ (e.g., the perfect transcript),
\begin{equation}
f(s^{t}) \le f_\text{max}, \quad \forall t,
\end{equation}
it follows from the Monotone Convergence Theorem that the sequence $\{ f(s^{t}) \}$ converges:
\begin{equation}
\lim_{t \to \infty} f(s^{t}) = f^* \le f_\text{max}.
\end{equation}
Therefore, the iterative optimization is guaranteed to converge to a transcript whose score cannot be further improved within the explored neighborhood.

\section{Experiments}\label{sec:experiments}
\subsection{Setting}
To evaluate the correction performance of the proposed method, we conduct experiments on the test subset of FLEURS dataset~\cite{conneau2022fleurs}, which covers 102 languages and contains 945 Chinese samples and 647 English samples. To examine its robustness across different ASR systems, we adopt Whisper-medium and Whisper-large-v3 as the base recognizers. Furthermore, to assess its adaptability to different correction models, Qwen3-235B~\cite{yang2025qwen3} and DeepSeek-V3.1~\cite{deepseekai2025deepseekv3} are employed. In all experiments, the maximum number of iterations is set to 8, and the candidate-pool size is fixed at 3. 

The baselines in our experiments include direct prompt~\cite{sachdev2024evolutionarypromptasr}, evolutionary prompt~\cite{sachdev2024evolutionarypromptasr}, and n-best~\cite{ma2023nbest}. To evaluate the contributions of individual components in our proposed method, we conduct an ablation study on DeepSeek-V3.1-based LIR-ASR by systematically removing rule constraints, multi-candidate selection, FSM, and neigbor searching. CER and WER are employed as evaluation metrics to quantify ASR correction performance.

\subsection{Main Result}
Table~\ref{tab:result} presents a comprehensive comparison of ASR correction methods applied to Whisper-medium and Whisper-large-v3 models in both Chinese and English. The proposed method LIR-ASR achieves the lowest CER/WER across all settings. 
Among the evaluated LLMs, DeepSeek-V3.1 yielded the most significant improvements when combined with LIR-ASR, achieving a CER of 3.20 and WER of 5.91 for Chinese, and CER of 1.68 and WER of 3.70 for English on the Whisper-medium model. This represents an average improvement of +0.86 in CER and +2.00 in WER over the baseline.
Notably, the Whisper-medium model with LIR-ASR correction achieved better performance than the Whisper-large-v3 baseline. Specifically, for Chinese, LIR-ASR on Whisper-medium achieved a CER of 3.20, surpassing the Whisper-large-v3 baseline CER of 3.98. For English, LIR-ASR on Whisper-medium achieved a WER of 3.70, surpassing the Whisper-large-v3 baseline WER of 4.23.
LIR-ASR demonstrated the highest average improvement across all tested configurations. The average improvements over the baseline were +0.88/+1.68 with Qwen3-235B and +0.86/+2.00 with Deepseek-V3.1, indicating the effectiveness of the proposed method in enhancing ASR correction performance.
Although evolutionary prompt aims to optimize the prompt iteratively, it lacks explicit linguistic or semantic constraints. As a result, in most scenarios, it performs worse than the uncorrected baseline.
The n-best method consistently performs better than both direct and evolutionary prompts, serving as a strong baseline. Increasing the number of candidates from 3-best to 6-best further improves correction performance, as seen with Qwen3-235B on Whisper-medium: the CER for Chinese decreases from 4.80 (3-best) to 4.54 (6-best). This demonstrates that expanding the candidate pool allows the system to select more accurate alternatives.

\begin{figure}[t!]
  \includegraphics[width=\columnwidth]{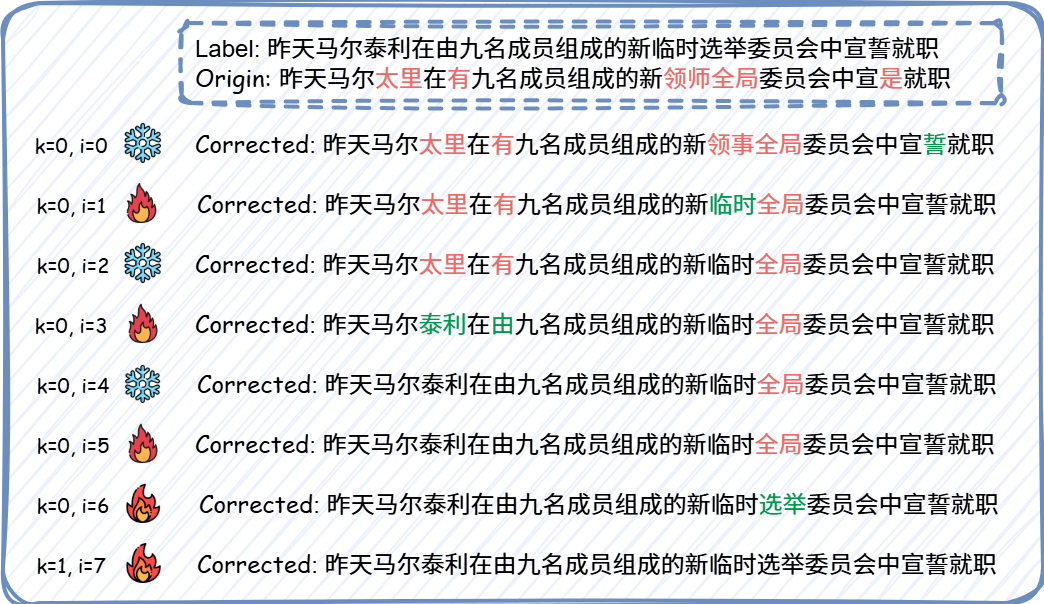}
  \caption{An example of ASR correction.}
  \label{fig:example}
\end{figure}

\subsection{Ablation study}
The ablation results are shown in the last four rows of Table~\ref{tab:result}. Removing rule-based constraints caused the largest performance drop, e.g., CER/WER on Whisper-large-v3 and DeepSeek-V3.1 for Chinese increased from 2.89/5.23 to 6.62/9.27, emphasizing their critical role in guiding the LLM to avoid linguistically plausible but semantically inconsistent substitutions. Omitting multi-candidate generation led to a rise in CER/WER to 3.24/6.40, confirming that generating diverse candidate variants is essential for effective iterative refinement and improving transcription quality. Excluding the FSM slightly degraded performance to 3.52/6.46, showing that state-based control helps the heuristic optimization process avoid local optima and better manage interdependent recognition errors. Finally, removing the neighbor searching mechanism increased CER/WER to 3.25/5.98, demonstrating that controlled randomness during the “Imagining” step enables exploration of a wider solution space, supporting gradual convergence to more accurate corrections.

\subsection{Convergence Analysis}
\begin{figure}[t!]
  \centering
  \begin{subfigure}[t]{0.495\columnwidth}
    \centering
    \includegraphics[width=\linewidth]{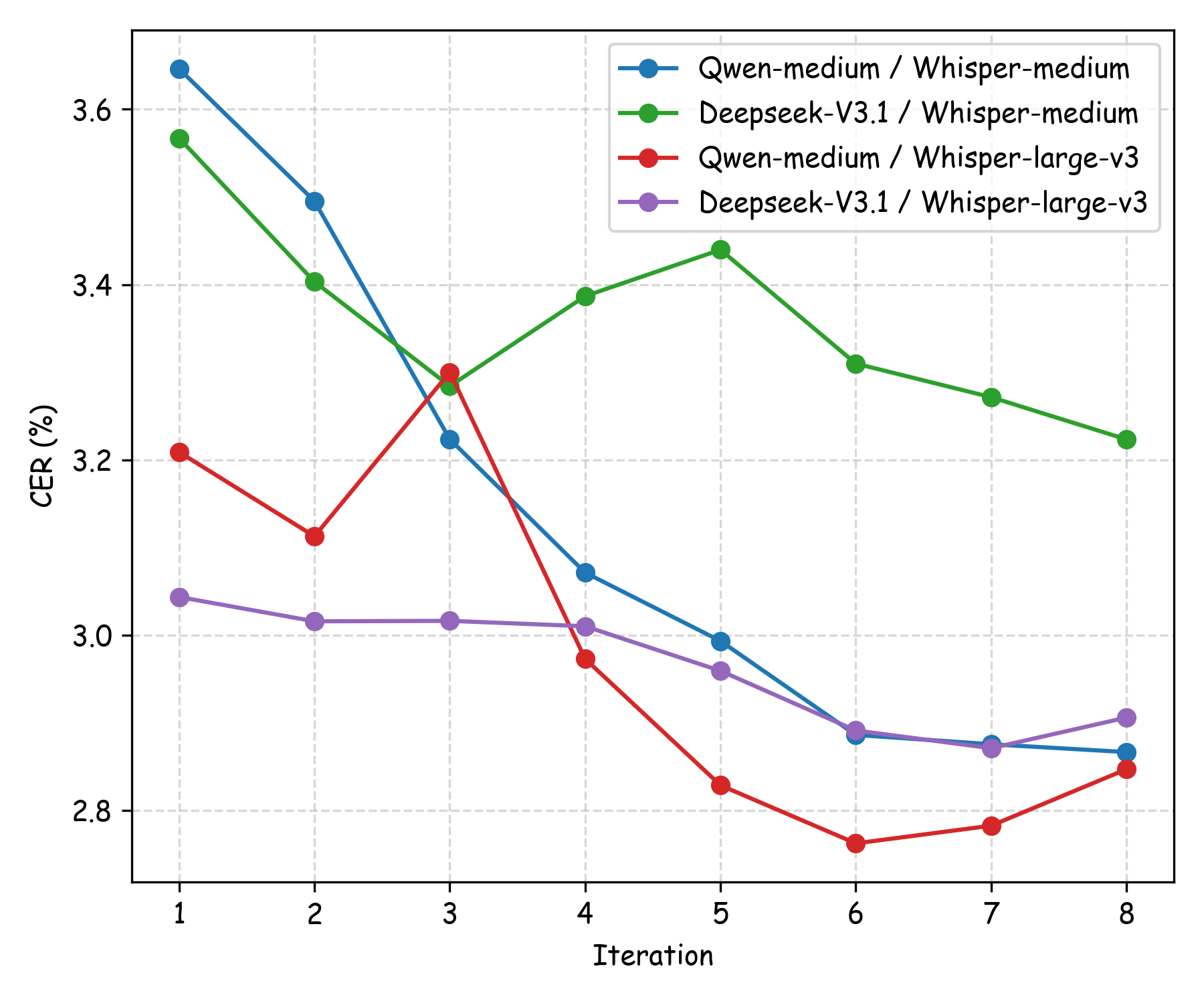}
    \caption{CER vs Iteration}
  \end{subfigure}
  \hfill
  \begin{subfigure}[t]{0.495\columnwidth}
    \centering
    \includegraphics[width=\linewidth]{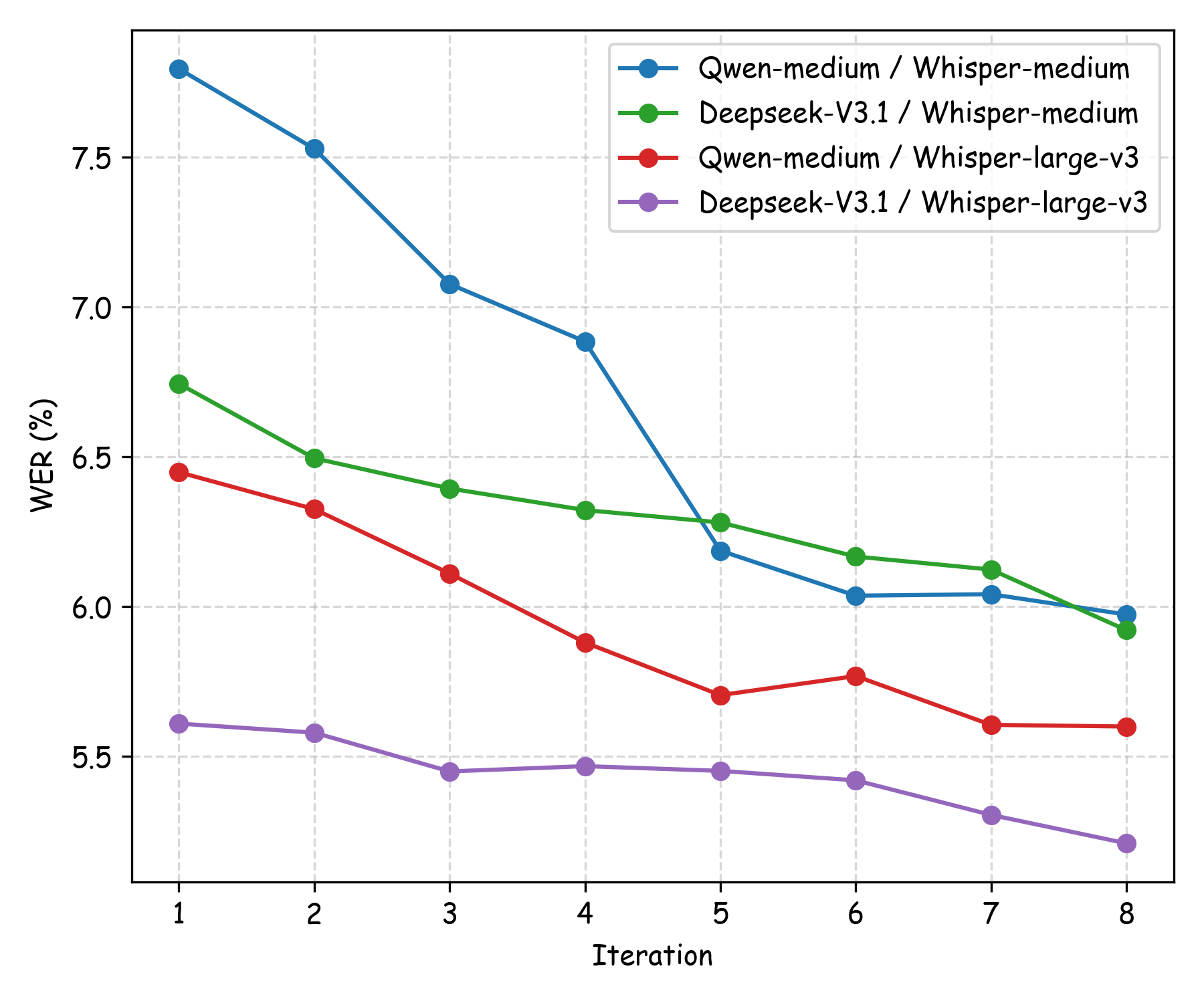}
    \caption{WER vs Iteration}
  \end{subfigure}
  \caption{Convergence visualization of the proposed LIR-ASR}
  \label{fig:convergence}
\end{figure}
To evaluate the effectiveness of iterative optimization in the proposed LIR-ASR framework, an illustrative example in Chinese is shown in Fig.~\ref{fig:example}. As depicted, ASR errors are progressively corrected through successive iterations. Obvious errors are addressed during the No Search and Search stages, while more subtle, nearly imperceptible errors are resolved in the Search++ stage.

To further assess the convergence behavior of our method, we tracked the WER and CER metrics of the transcripts generated at each iteration. Fig.~\ref{fig:convergence} presents the WER and CER curves over iterations for representative Chinese samples. The curves exhibit a monotonically non-increasing trend, indicating that the heuristic optimization consistently improves transcript quality. Additionally, the improvements gradually plateau after a few iterations, demonstrating that the method converges to a stable solution within the explored candidate neighborhood. These results confirm that our iterative heuristic optimization not only effectively enhances ASR outputs but also shows reliable and stable convergence behavior across different languages and ASR backbones.

\section{Conclusion}\label{sec:conclusion}
In this paper, we proposed \textbf{LIR-ASR}, a novel simulated heuristic optimized iterative ASR correction framework, leveraging large language models. 
Inspired by human auditory perception, LIR-ASR adopts a "Listening-Imagining-Refining" paradigm, in which uncertain words are first identified, then phonetically plausible alternatives are generated, and finally refined within the contextual constraints of the transcript.
By integrating a finite state machine to control the heuristic searching process and applying rule-based constraints to guide LLM scoring, our approach effectively mitigates contextually plausible recognition errors and reduces the risk of semantically inconsistent substitutions. Experiments on both English and Chinese ASR outputs show that LIR-ASR achieves average reductions
in CER/WER of up to 1.5 percentage points compared to baselines, demonstrating substantial accuracy gains in transcription. In future work, we plan to extend LIR-ASR to Tibetan and other low-resource languages, where phonetic variability and data scarcity pose additional challenges, further validating its cross-lingual adaptability.

\vfill\pagebreak
\section*{Acknowledgments}
This work was supported in part by the \emph{National Natural Science Foundation of China under Grants 62276055 and 62406062},in part by the \emph{Sichuan Science and Technology Program under Grant 2023YFG0288}, in part by the \emph{Natural Science Foundation of Sichuan Province under Grant 2024NSFSC1476},in part by the \emph{National Science and Technology Major Project under Grant 2022ZD0116100},in part by the \emph{Sichuan Provincial Major Science and Technology Project under Grant 2024ZDZX0012}.

\bibliographystyle{IEEEbib}
\bibliography{strings,refs,ref_asr,ref_correction,ref_llm}

\end{document}